# A Distributed Probabilistic Modeling Algorithm for the Aggregated Power Forecast Error of Multiple Newly Built Wind Farms

Mengshuo Jia, *Student Member, IEEE*, Chen Shen, *Senior Member, IEEE*, Zhiwen Wang, *Student Member, IEEE*

*Abstract*—The extensive penetration of wind farms (WFs) presents challenges to the operation of distribution networks (DNs). Building a probability distribution of the aggregated wind power forecast error is of great value for decision making. However, as a result of recent govern -ment incentives, many WFs are being newly built with little historical data for training distribution models. Moreover, WFs with different stakeholders may refuse to submit the raw data to a data center for model training. To address these problems, a Gaussian mixture model (GMM) is applied to build the distribution of the aggregated wind power forecast error; then, the maximum a posteriori (MAP) estimation method is adopted to overcome the limited training data problem in GMM parameter estimation. Next, a distributed MAP estimation method is developed based on the average consensus filter algorithm to address the data privacy issue. The distribution control center is introduced into the distributed estimation process to acquire more precise estimation results and better adapt to the DN control architecture. The effectiveness of the proposed algorithm is empirically verified using historical data.

*Index Terms*: Wind farms, probabilistic modeling, limited training data, data privacy, maximum a posteriori estimation, distributed parametric estimation

## I. INTRODUCTION

WIND generation (WG) is an economical and environmentally friendly mean of producing electricity and is increasingly implemented worldwide. A WG adoption level of 20% in the U.S. is expected by the year 2030 [1]. Moreover, the penetration of wind farms (WFs) in distributed networks (DNs) has greatly increased [2]. In China, the integration of WFs into DNs is encouraged in the 13th Five-year Plan of energy development [3]. However, extensive penetration of WFs raises challenges to the operation of DNs because of the variability of the wind speed [4], [5]. As an effective method to quantitatively evaluate the impact of this uncertainty, constructing a conditional probability distribution function (PDF) of the aggregated wind power forecast error (WFE) for a given forecast wind power output (FWO) in a DN is of great importance for optimal decision making [6], [7].

As the most commonly used probability distribution for characterizing wind power uncertainty, the Weibull distribu -tion is also used to model the PDF of WFE [8]. However, the Cauchy distribution fits better than the Weibull distribution at the hour and 15-minute timescales [8]. Except for the Cauchy distribution, Gaussian distribution [9], [10] and beta distribution [11] have also been utilized to construct the PDF of WFE. However, these distributions cannot be applied at all different time scales [12]. To reflect the skewed and heavy-tailed features of the uncertainty, the α-stable distribution [13], the mixed beta distribution [14] and the versatile probability distribution model [15] have been proposed to improve the fitting performance. Known for the good performance in high-precision fitting, the Gaussian mixture model (GMM) can characterize random variables that obey arbitrary distributions. Thus, for probabilistic analysis and stochastic optimizations in a power system, GMMs have been used to formulate the PDF of wind power uncertainty [16]–[18]. In [16], the cumulative distribution function of the wind power forecast errors is constructed with a GMM. In [17], the PDF of the actual wind power output (AWO) is formulated by a customized GMM. In [18], a GMM is adopted to construct the joint PDF of AWO for multiple wind farms. These studies not only show the high precision of GMM in describing the non-Gaussian uncertainty of the WFs but also show the convenience of the use of GMM-based PDF in function transformations. Therefore, GMM is also applied in this paper to formulate the PDF of wind power uncertainty

The essence of constructing a PDF using a GMM lies in estimating the parameters of the model. As a maximum likelihood (ML) estimation, the expectation maximization (EM) algorithm is commonly used for GMM [16]–[18]. Many commercial software tools provide reliable off-the-shelf solvers for the EM algorithm [19]. As long as there is sufficient historical data for training, the EM algorithm can calculate the parameters of GMM with high accuracy.

However, in response to the recent government incentives, many WFs are rapidly being built. These newly built WFs (NWFs) lack historical data. With limited training data, the effectiveness of the EM algorithm is weakened[20], resulting in overfitting [21], [22]. To solve the limited training data problem, Bayesian estimation is a feasible approach that is

This work was supported in part by the Foundation for Innovative Research Groups of the National Natural Science Foundation of China under Grant 51621065 and in part by the Joint Funds of the National Natural Science Foundation of China under Grant U1766206 (*Corresponding to Chen Shen*).

M. Jia, C. Shen, and Z. Wang are with the State Key Laboratory of Power Systems, Department of Electrical Engineering, Tsinghua University, Beijing 100084, China (e-mails: jms16@mails.tsinghua.edu.cn, shenchen@mail.tsinghua.edu.cn, wang-zw13@mails.tsinghua.edu.cn).

already used in the speech and image recognition field [20]–[25]. Bayesian estimation can be divided into two categories: point estimation and predictive estimation. The former can be realized by maximum a posteriori (MAP) estimation [23], and the latter requires the use of the variational learning method [24] or the Monte Carlo Markov Chain (MCMC) method [26]. Because the MAP estimation does not require high dimen-sional complex integrals, it greatly reduces the computation burden. Thus, MAP estimation is utilized in this paper to estimate the parameters of GMM when the training data are not sufficient.

To construct the conditional PDF of the aggregated WFE in a DN for a given aggregated FWO, the joint PDF of the aggregated AWO and FWO must first be constructed. Then, the conditional PDF can be derived from the joint PDF using simple algebraic calculation [19]. For constructing the joint PDF, common MAP estimation methods must gather the raw data pairs of AWO and FWO from all the NWFs to form the aggregated AWO and FWO historical data for training. However, NWFs with different stakeholders may refuse to submit raw data to a data center for the protection of data privacy. Moreover, the centralized communication structure may suffer from single point failure if the data center is down for any reason. In the field of wireless sensor networks, the distributed parameter estimation algorithm with distributed communication structure has been proposed [27], [28]. In [27], sensors are connected in a circle and pass information clockwise. However, when the network becomes complex, the communication will be time consuming. In [28], an average consensus filter (ACF) algorithm is presented, in which sensors only need to communicate with their neighbors to exchange processed data instead of raw data. Moreover, this algorithm enables every sensor to calculate the global distribution. Thus, the ACF algorithm is adopted in this paper, which ensures each NWF in a DN to build the joint PDF of the aggregated AWO and FWO. Then, the conditional PDF of the aggregated WFE can be derived from the joint PDF by each NWF via algebraic calculation [19]. Of course, only one estimated result is required in the distributed control center (DCC) of DN to make an optimal decision. Thus, DCC must choose one estimated result that best matches the real distribution.

Based on GMM, this paper focuses on investigating a distributed MAP algorithm to construct the joint PDF of the aggregated AWO and FWO in a DN. The original contributions of this paper are as follows:
- The MAP estimation is applied for estimating the parameters of GMM-based joint PDF of the aggregated AWO and FWO in a DN to solve the limited training data problem.
- A distributed MAP (DMAP) estimation algorithm is developed to protect the data privacy of NWFs.
- DCC is introduced into the proposed DMAP estimation algorithm as a virtual node (VN) to select the best constructed distribution for decision making and better fit the DN control paradigm. Moreover, the proposed algorithm is robust, and can deal with single point failure.

Notably, since the joint PDF is the basis of the conditional PDF, the main concern of this paper is building the joint PDF accurately while considering the limited training data problem and the data privacy problem. However, for the integrity of this paper, construction of the conditional PDF of the aggregated WFE based on the joint PDF of the aggregated AWO and FWO is also provided, and the case study of the constructed conditional PDF is illustrated as well.

The rest of the paper is organized as follows. In Section II, GMM-based distributions, including the joint PDF, marginal PDF and the conditional PDF, are introduced. In Section III, the centralized MAP estimation is detailed. In Section IV, the distributed MAP estimation is proposed. In Section V, DCC is introduced into the distributed MAP estimation. Case studies are described in Section VI. Finally, Section VII concludes this paper.

## II. GMM-BASED DISTRIBUTIONS

In this section, GMM-based joint PDF of the aggregated AWO and FWO, GMM-based marginal PDFs of the aggregated AWO and FWO, and GMM-based conditional PDF of the aggregated WFE are briefly introduced.

### A. The joint PDF of the aggregated AWO and FWO

Assume that there are $M$ NWFs integrated into a DN. Let the subscript $m$ be the index of NWFs. The random variables of AWO and FWO are denoted by $\mathbf{Y}_m=[\mathbf{y}_m^A, \mathbf{y}_m^F]$ ($m=1,\ldots, M$) where $\mathbf{y}_m^A$ denotes AWO and $\mathbf{y}_m^F$ denotes FWO. The random variables of the aggregated AWO and FWO are defined as $\mathbf{Z}$ as follows:

$$\mathbf{Z} = \begin{bmatrix} \mathbf{Z}^A & \mathbf{Z}^F \end{bmatrix} = \begin{bmatrix} \sum_{m=1}^{M} \mathbf{y}_m^A & \sum_{m=1}^{M} \mathbf{y}_m^F \end{bmatrix} \quad (1)$$

$\mathbf{Y}_m$ has $N$ output observations $\mathbf{y}_{m,i}$ ($m=1,\ldots, M$, $i=1,\ldots, N$). Thus, $\mathbf{Z}$ also has $N$ aggregated output observations $\mathbf{z}_i$ ($i=1,\ldots, N$) as follows:

$$\mathbf{y}_{m,i} = \begin{bmatrix} y_{m,i}^A & y_{m,i}^F \end{bmatrix}, m=1,...,M, i=1,...,N, \quad (2)$$

$$\mathbf{z}_i = \begin{bmatrix} z_i^A & z_i^F \end{bmatrix} = \begin{bmatrix} \sum_{m=1}^{M} y_{m,i}^A & \sum_{m=1}^{M} y_{m,i}^F \end{bmatrix}, i=1,...,N \quad (3)$$

GMM is a parametric model represented by a weighted sum of Gaussian component densities [16]. We consider a GMM with $J$ Gaussian components. The weighted coefficient is denoted by $w_j$, and the $j$th Gaussian component is represented by mean $\boldsymbol{\mu}_j$ and covariance $\boldsymbol{\Sigma}_j$. Let $\boldsymbol{\theta}=\{w_j, \boldsymbol{\mu}_j, \boldsymbol{\Sigma}_j \mid j=1,\ldots, J \}$ be a parameter set of GMM. Thus, the joint PDF of the aggregated AWO and FWO in GMM is specified by (4), where $\boldsymbol{\mu}_j$ and $\boldsymbol{\Sigma}_j$ are specified via (5).

$$f(\mathbf{Z}|\boldsymbol{\theta}) = \sum_{j=1}^{J} w_j N(\mathbf{Z}|\boldsymbol{\mu}_j, \boldsymbol{\Sigma}_j)$$

$$N(\mathbf{Z}|\boldsymbol{\mu}_j, \boldsymbol{\Sigma}_j) = \frac{1}{(2\pi) \times \det(\boldsymbol{\Sigma}_j)^{1/2}}$$

$$\times \exp\left[-\frac{1}{2}(\mathbf{Z}-\boldsymbol{\mu}_j)^T \boldsymbol{\Sigma}_j^{-1}(\mathbf{Z}-\boldsymbol{\mu}_j)\right] \quad (4)$$

$$\boldsymbol{\mu}_j = \begin{bmatrix} \mu_j^A \\ \mu_j^F \end{bmatrix}, \boldsymbol{\Sigma}_j = \begin{bmatrix} \Sigma_j^{A,A} & \Sigma_j^{A,F} \\ \Sigma_j^{F,A} & \Sigma_j^{F,F} \end{bmatrix} \quad (5)$$

### B. The marginal PDFs of the aggregated AWO and FWO

Once the joint PDF in (4) is obtained, the marginal PDFs of the aggregated AWO and FWO can be directly calculated by (6).

$$f(\mathbf{Z}^A) = \sum_{j=1}^{J} w_j N(\mathbf{Z}^A; \mu_j^A, \Sigma_j^{A,A})$$
$$f(\mathbf{Z}^F) = \sum_{j=1}^{J} w_j N(\mathbf{Z}^F; \mu_j^F, \Sigma_j^{F,F}) \quad (6)$$

### C. The conditional PDF of the aggregated WFE

We define the variable of the aggregated WFE as $\mathbf{Z}^E = \mathbf{Z}^A - \mathbf{Z}^F$ and denote the given aggregated FWO by $z^F$. Once the joint PDF in (4) is obtained, the conditional PDF of the aggregated WFE for $z^F$ in (7) can be directly calculated via (8) [19].

Obviously, the joint PDF is the basis for building the conditional PDF. Meanwhile, as long as the joint PDF is accurate, the corresponding conditional PDF will also be accurate. Therefore, considering the limited training data problem and data privacy issue, the problem, '*how to build the conditional PDF of the aggregated WFE,*' changes into, '*how to build the joint PDF of aggregated AWO and FWO.*' The latter problem is the main concern of this paper.

$$f(\mathbf{Z}^E | z^F) = \sum_{j=1}^{J} w_j^c N(\mathbf{Z}^E + z^F; \mu_j^c, \Sigma_j^c) \quad (7)$$

$$\begin{cases} w_j^c = w_j \dfrac{N(\mathbf{Z}^F; \mu_j^F, \Sigma_j^{F,F})}{\sum_{j=1}^{J} w_j N(\mathbf{Z}^F; \mu_j^F, \Sigma_j^{F,F})} \\ \mu_j^c = \mu_j^A + \Sigma_j^{A,F} \times (\Sigma_j^{F,F})^{-1} \times (\mathbf{Z}^F - \mu_j^{F,F}) \\ \Sigma_j^c = \Sigma_j^{A,A} - \Sigma_j^{A,F} \times (\Sigma_j^{F,F})^{-1} \times \Sigma_j^{F,A} \end{cases} \quad (8)$$

## III. CENTRALIZED MAP ESTIMATION FOR THE JOINT PDF

The centralized MAP estimation for constructing the joint PDF of the aggregated AWO and FWO is introduced first, laying a foundation for developing the distributed MAP estimation. As a Bayesian estimation approach, the MAP estimation considers the values of $\boldsymbol{\theta}$ as variables instead of constants. The distribution of $\boldsymbol{\theta}$ is defined as the prior distribution $\pi(\boldsymbol{\theta})$. The key of the MAP estimation is to maximize the posteriori distribution $\pi(\boldsymbol{\theta}|\mathbf{Z})$ to obtain $\boldsymbol{\theta}_{MAP}$:

$$\boldsymbol{\theta}_{MAP} = \operatorname*{argmax}_{\boldsymbol{\theta}} \pi(\boldsymbol{\theta}|\mathbf{Z}) = \operatorname*{argmax}_{\boldsymbol{\theta}} f(\mathbf{Z}|\boldsymbol{\theta})\pi(\boldsymbol{\theta}) \quad (9).$$

Solving (9) first requires two conditions: one is to obtain the aggregated output $\mathbf{Z}$, and the other is to obtain the prior distribution $\pi(\boldsymbol{\theta})$.

For the aggregated output $\mathbf{Z}$, the centralized MAP estimation requires the raw data pairs $y_{m,i}$ of all the NWFs to be sent to a data center. Then, the aggregated output $z_i$ can be obtained using (1).

For the prior distribution, based on the current studies on GMM, the following priors are considered as practical candidates [20], [24], [26]: a joint Dirichlet distribution for the weighted coefficients, as in (10), and a joint Normal-Wishart distribution for the $j$th mean vectors and the inverse of the $j$th covariance matrix, as in (11).

$$h(w_1,...,w_J | v_1,...,v_J) \propto \prod_{j=1}^{J} w_j^{v_j - 1} \quad (10)$$

$$h(\boldsymbol{\mu}_j, \boldsymbol{\rho}_j | \boldsymbol{\lambda}_j, \tau_j, \alpha_j, \boldsymbol{\sigma}_j) \propto |\boldsymbol{\rho}_j|^{(\alpha_j - d)/2}$$
$$\times \exp\left[-\frac{\tau_j}{2}(\boldsymbol{\mu}_j - \boldsymbol{\lambda}_j)^T \boldsymbol{\rho}_j (\boldsymbol{\mu}_j - \boldsymbol{\lambda}_j)\right] \times \exp\left[-\frac{1}{2} tr(\boldsymbol{\sigma}_j \boldsymbol{\rho}_j)\right] \quad (11)$$

where $v_j > 0$ represents the hyperparameters for the Dirichlet distribution. $\boldsymbol{\lambda}_j$, $\tau_j$, $\alpha_j$ and $\boldsymbol{\sigma}_j$ are the hyperparameters for the Normal-Wishart distribution. $tr(\cdot)$ is the trace function. $\boldsymbol{\rho}_j$ is the inverse of the covariance $\boldsymbol{\Sigma}_j$, named the precision matrix. Let $\boldsymbol{\varphi} = \{w_j, \boldsymbol{\mu}_j, \boldsymbol{\rho}_j \mid j=1,\ldots,J\}$ be the parameter set after the introduction of the precision matrix $\boldsymbol{\rho}_j$. Note that $\boldsymbol{\varphi}$ and $\boldsymbol{\theta}$ have a unique correspondence. Once $\boldsymbol{\varphi}$ is known, $\boldsymbol{\theta}$ can be obtained through the inverse calculation of $\boldsymbol{\rho}_j$.

The joint prior distribution of $\boldsymbol{\varphi}$ is the product of the prior distributions defined in (10) and (11) as follows:

$$\pi(\boldsymbol{\varphi}) = h(w_1,...,w_J) \times \prod_{j=1}^{J} h(\boldsymbol{\mu}_j, \boldsymbol{\rho}_j) \quad (12)$$

Once the aggregated output $\mathbf{Z}$ and the prior distribution $\pi(\boldsymbol{\varphi})$ are obtained, (9) can be modified to (13) based on the unique correspondence between $\boldsymbol{\varphi}$ and $\boldsymbol{\theta}$ as follows:

$$\boldsymbol{\varphi}_{MAP} = \operatorname*{argmax}_{\boldsymbol{\varphi}} f(\mathbf{Z}|\boldsymbol{\theta})\pi(\boldsymbol{\varphi}) \quad (13)$$

Equation (13) can be solved by a two-step iterative calculation [23]. In the first step of the $(t+1)$th iteration, the following statistics of the $j$th Gaussian component are calculated with the aggregated output $z_i$ and parameters $\boldsymbol{\varphi} = \{w_j, \boldsymbol{\mu}_j, \boldsymbol{\rho}_j \mid j=1,\ldots,J\}$ estimated in the $t$th iteration as follows:

$$C_{j,i}^{t+1} = \frac{w_j^t N(\mathbf{z}_i; \boldsymbol{\mu}_j^t, \boldsymbol{\rho}_j^t)}{\sum_{k=1}^{J} w_k^t N(\mathbf{z}_i; \boldsymbol{\mu}_k^t, \boldsymbol{\rho}_k^t)}, i=1,...,N$$

$$C_j^{t+1} = \sum_{i=1}^{N} C_{j,i}^t, \quad \boldsymbol{\chi}_j^{t+1} = \frac{\sum_{i=1}^{N} C_{j,i}^t \mathbf{z}_i}{\sum_{i=1}^{N} C_{j,i}^t} \quad (14)$$

$$\boldsymbol{\psi}_j^{t+1} = \sum_{i=1}^{N} C_{j,i}^t (\mathbf{z}_i - \boldsymbol{\chi}_j^{t+1})(\mathbf{z}_i - \boldsymbol{\chi}_j^{t+1})^T$$

In the second step of the $(t+1)$th iteration, parameters $\boldsymbol{\varphi} = \{w_j, \boldsymbol{\mu}_j, \boldsymbol{\rho}_j \mid j=1,\ldots,J\}$ are updated using the hyperparameters and the statistics calculated in (14) as follows:

$$w_j^{t+1} = \frac{v_j + C_j^{t+1} - 1}{\sum_{j=1}^{J}(v_j + C_j^{t+1} - 1)}$$

$$\boldsymbol{\mu}_j^{t+1} = \frac{\tau_j \boldsymbol{\lambda}_j + C_j^{t+1} \boldsymbol{\chi}_j^{t+1}}{\tau_j + C_j^{t+1}} \quad (15)$$

$$\boldsymbol{\rho}_j^{t+1} = \frac{\alpha_j + C_j^{t+1} - d}{\boldsymbol{\sigma}_j + \boldsymbol{\psi}_j^{t+1} + \dfrac{\tau_j C_j^{t+1}}{\tau_j + C_j^{t+1}}(\boldsymbol{\lambda}_j - \boldsymbol{\chi}_j^{t+1})(\boldsymbol{\lambda}_j - \boldsymbol{\chi}_j^{t+1})^T}$$

These two steps are calculated iteratively until convergence is achieved.

Although the centralized MAP estimation is an easy method to obtain the joint PDF of the aggregated AWO and FWO, this algorithm requires that the raw data of all NWFs be gathered in a data center. Such centralized communication puts data privacy at risk. Moreover, this algorithm suffers from single point failure if the data center fails.

## IV. DISTRIBUTED MAP ESTIMATION FOR THE JOINT PDF

Centralized data processing is only required in (14), which is the first step of the centralized MAP estimation. The calculation of the second step in (15) does not require the raw data but depends solely on the statistics calculated by (14). Therefore, to develop a DMAP estimation, the key is to decentralize the first step of the centralized MAP estimation.

In this section, we propose a DMAP estimation algorithm. This algorithm has two stages. First, each NWF estimates the aggregated output of all NWFs. Second, each of the estimated aggregated outputs is trained in a distributed manner to construct the joint PDF of the aggregated AWO and FWO.

### A. Estimating the aggregated output

To avoid gathering the raw data from all the NWFs, we first enable every NWF to estimate the aggregated output using the ACF algorithm [28]. The estimated result of the $m$-th NWF, named the estimated aggregated output in this paper, is denoted by $\mathbf{z}_{m,i}$ ($m=1,…,M$, $i=1,…,N$).

Certain definitions are given first: the $M$ NWFs are viewed as $M$ nodes in a communication network, as shown in Fig. 1. Any pair of nodes is connected if the distance between the two nodes is less than a threshold. The threshold should guarantee that the whole network is connected. The adjacent nodes of each node are defined as its one-hop neighbors, and each node only communicates with its adjacent nodes. The adjacent domain of the $m$-th node is represented by $\Omega_m$ and is given in the dotted box in Fig. 1.

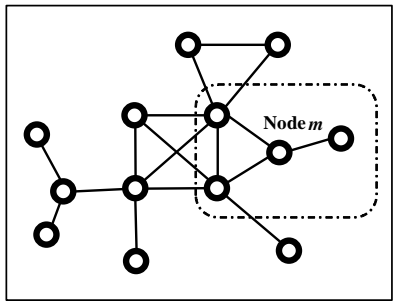

Fig. 1 The schematic diagram of the communication network

Each NWF first estimates the average output of all NWFs that is represented by $\bar{\mathbf{z}}_{m,i}$ ($m=1,…, M$, $i=1,…, N$) through iteration using the ACF algorithm. For the ($t_c+1$)th iteration, $\bar{\mathbf{z}}_{m,i}$ is denoted by $\bar{\mathbf{z}}_{m,i}^{t_c+1}$ and calculated via (16) as follows:

$$\bar{\mathbf{z}}_{m,i}^{t_c+1} = \bar{\mathbf{z}}_{m,i}^{t_c} + \eta\left[\mathbf{y}_{m,i} - \bar{\mathbf{z}}_{m,i}^{t_c} + \sum_{n\in\Omega_m}\left(\bar{\mathbf{z}}_{n,i}^{t_c} - \bar{\mathbf{z}}_{m,i}^{t_c}\right)\right], i=1,...N \quad (16)$$

where $n$ is the index of the NWF that belongs to the adjacent domain of the $m$-th NWF, and $\bar{\mathbf{z}}_{n,i}^{t_c}$ is the estimated average output calculated by the $n$-th NWF in the $t_c$th iteration. $\eta$ is a predetermined updating rate.

According to the LaSalle invariance principle, $\bar{\mathbf{z}}_{m,i}$ in (16) is global asymptotically ε stable to the real average output of all NWFs [28]. Thus, multiplying the converged $\bar{\mathbf{z}}_{m,i}$ by the number of NWFs, which is $M$, allows $\mathbf{z}_{m,i}$ to be obtained for the $m$-th NWF:

$$\mathbf{z}_{m,i} = M \times \bar{\mathbf{z}}_{m,i}, \ i=1,...,N \quad (17)$$

Finally, each NWF obtains its $\mathbf{z}_{m,i}$ ($i=1,…, N$).

In the above process, communications only occur in (16). When the $m$-th NWF calculates its $\bar{\mathbf{z}}_{m,i}$, only $\bar{\mathbf{z}}_{n,i}$ from its adjacent NWFs is required. Thus, the exchanged data is $\bar{\mathbf{z}}_{n,i}$ instead of the raw data, and the data privacy is protected.

### B. Constructing the joint PDF

Once the estimated aggregated output is obtained, each NWF can directly construct the joint PDF of the aggregated AWO and FWO via (14)-(15) by replacing $\mathbf{z}_i$ with its own $\mathbf{z}_{m,i}$ ($i=1,…, N$). However, since the calculation of the estimated aggregated output only depends on the adjacent NWFs, the estimated aggregated outputs are not exactly equal to the real values. Moreover, different NWFs have different adjacent domains, thus the gap between the estimated aggregated output and the real value varies when the NWF varies. Therefore, directly constructing the joint PDF by only utilizing the estimated aggregated output of one NWF cannot achieve the desired accuracy. The simulation results are verified in Section VI.

After the estimated aggregated output is obtained, each NWF can be considered as a sensor that observes the aggregated output in a DN. The observation of the $m$-th sensor is its $\mathbf{z}_{m,i}$ ($i=1,…, N$). To construct a joint PDF that better describes the real distribution of the aggregated output in a DN, the observations of all the sensors should be collected and trained.

The training process is the two-step iterative calculation in (14) and (15). Considering the collection of all the observations, the first step of the ($t$+1)th iteration is modified as follows:

$$C_{j,m,i}^{t+1} = \frac{w_j^t N(\mathbf{z}_{m,i}; \boldsymbol{\mu}_j^t, \boldsymbol{\rho}_j^t)}{\sum_{k=1}^{J} w_k^t N(\mathbf{z}_i; \boldsymbol{\mu}_k^t, \boldsymbol{\rho}_k^t)}$$

$$C_j^{t+1} = \sum_{m=1}^{M}\sum_{i=1}^{N} C_{j,m,i}^t$$

$$\boldsymbol{\chi}_j^{t+1} = \sum_{m=1}^{M} \frac{\sum_{i=1}^{N} C_{j,m,i}^t \mathbf{z}_{m,i}}{\sum_{i=1}^{N} C_{j,m,i}^t} \quad (18)$$

$$\boldsymbol{\psi}_j^{t+1} = \sum_{m=1}^{M}\sum_{i=1}^{N} C_{j,m,i}^t (\mathbf{z}_{m,i} - \boldsymbol{\chi}_j^{t+1})(\mathbf{z}_{m,i} - \boldsymbol{\chi}_j^{t+1})^T$$

To calculate (18) in a distributed manner, we first define the local statistics $\mathbf{L}_{j,m} = \{ C_{j,m}, \boldsymbol{\beta}_{j,m}, \boldsymbol{\gamma}_{j,m}, \mathbf{S}_{j,m}, \boldsymbol{\chi}_{j,m} \}$ for the $j$th component and the $m$-th NWF as follows:

$$C_{j,m}^t = \sum_{i=1}^{N} C_{j,m,i}^t, \quad \boldsymbol{\beta}_{j,m}^t = \sum_{i=1}^{N} C_{j,m,i}^t \mathbf{z}_{m,i}$$

$$\boldsymbol{\gamma}_{j,m}^t = \sum_{i=1}^{N} C_{j,m,i}^t \mathbf{z}_{m,i}^T, \quad \mathbf{S}_{j,m}^t = \sum_{i=1}^{N} C_{j,m,i}^t \mathbf{z}_{m,i} \mathbf{z}_{m,i}^T \quad (19)$$

$$\boldsymbol{\chi}_{j,m}^t = \boldsymbol{\beta}_{j,m}^t / C_{j,m}^t$$

Note that each NWF can calculate its own local statistics using its own observations $\mathbf{z}_{m,i}$ ($i=1,\ldots,N$) without communicating with its neighbors.

Then, based on the local statistics, (18) can be reformulated into (20) as follows:

$$C_j^{t+1} = \sum_{m=1}^{M} C_{j,m}^t, \quad \boldsymbol{\beta}_j^{t+1} = \sum_{m=1}^{M} \boldsymbol{\beta}_{j,m}^t$$

$$\boldsymbol{\gamma}_j^{t+1} = \sum_{m=1}^{M} \boldsymbol{\gamma}_{j,m}^t, \quad \mathbf{S}_j^{t+1} = \sum_{m=1}^{M} \mathbf{S}_{j,m}^t \quad (20)$$

$$\boldsymbol{\chi}_j^{t+1} = \sum_{m=1}^{M} \boldsymbol{\chi}_{j,m}^t$$

$$\boldsymbol{\psi}_j^{t+1} = \mathbf{S}_j^t + \boldsymbol{\chi}_j^t \left(\boldsymbol{\chi}_j^t\right)^T C_j^t - \boldsymbol{\beta}_j^t \left(\boldsymbol{\chi}_j^t\right)^T - \boldsymbol{\chi}_j^t \boldsymbol{\gamma}_j^t$$

The statistics $\mathbf{G}_j = \{\Sigma C_{j,m}, \Sigma \boldsymbol{\beta}_{j,m}, \Sigma \boldsymbol{\gamma}_{j,m}, \Sigma \mathbf{S}_{j,m}, \Sigma \boldsymbol{\chi}_{j,m}\}$ in (20) are defined as the global statistics. The relationship between the local statistics and the global statistics is as follows:

$$\mathbf{G}_j = \sum_{m=1}^{M} \mathbf{L}_{j,m} \quad (21)$$

The ACF algorithm can be utilized for each NWF to estimate the global statistics based on its local statistics and exchanged data with its neighbors. The estimated results, named estimated global statistics, are denoted by $\mathbf{G}_{j,m}$ ($m=1,\ldots,M$). The estimated average global statistics are denoted by $\overline{\mathbf{G}}_{j,m}$ ($m=1,\ldots,M$). For the ($t_c+1$)th iteration in the ACF algorithm, $\overline{\mathbf{G}}_{j,m}$ is represented by $\overline{\mathbf{G}}_{j,m}^{t_c+1}$ and calculated by (22):

$$\overline{\mathbf{G}}_{j,m}^{t_c+1} = \overline{\mathbf{G}}_{j,m}^{t_c} + \eta \left[ \mathbf{L}_{j,m} - \overline{\mathbf{G}}_{j,m}^{t_c} + \sum_{n \in \Omega_m} \left( \overline{\mathbf{G}}_{j,n}^{t_c} - \overline{\mathbf{G}}_{j,m}^{t_c} \right) \right] \quad (22)$$

where $n$ is the index of the NWF that belongs to the adjacent domain of the $m$th NWF, and $\overline{\mathbf{G}}_{j,n}^{t_c}$ is the estimated average global statistics calculated by the $n$-th NWF in the $t_c$th iteration. Once the converged $\overline{\mathbf{G}}_{j,m}$ is obtained, by multiplying it with $M$, the converged estimated global statistics $\mathbf{G}_{j,m}$ can be obtained as follows:

$$\mathbf{G}_{j,m} = M \times \overline{\mathbf{G}}_{j,m}, \quad m=1,\ldots,M, \ j=1,\ldots,J \quad (23)$$

After the converged $\mathbf{G}_{j,m}$ is obtained, each NWF can calculate $\boldsymbol{\Psi}_j$ in (20) independently and complete the first step of the ($t+1$)th iteration.

For the second step of the ($t+1$)th iteration in (15), each NWF can update its $\boldsymbol{\varphi} = \{w_j, \boldsymbol{\mu}_j, \boldsymbol{\rho}_j \mid j=1,\ldots,J\}$ directly using its estimated global statistics and $\boldsymbol{\Psi}_j$. Thereafter, the two steps are calculated iteratively until convergence, and joint PDFs will be constructed.

Note that because different NWFs will obtain different converged $\mathbf{G}_{j,m}$, the updated $\boldsymbol{\varphi} = \{w_j, \boldsymbol{\mu}_j, \boldsymbol{\rho}_j \mid j=1,\ldots,J\}$ will also differ from those of other NWFs. After the two-step iterative converges, $M$ NWFs will construct $M$ different joint PDFs of the aggregated AWO and FWO. Due to the characteristics of the ACF algorithm, the differences between these joint PDFs are very small.

In the above iteration, communications only occur in (22), and the exchange data is $\overline{\mathbf{G}}_{j,n}$ instead of raw data. Thus, the data privacy is protected.

## V. Distributed MAP Estimation With DCC

The proposed DMAP estimation in Section IV enables every NWF to obtain a joint PDF only using its observations and certain data exchanged with its neighbors. These PDFs are similar but not identical. For DCC to make an optimal decision, the most accurate joint PDF is the most desirable. Therefore, we introduce DCC into the proposed distributed MAP estimation to solve the problem of selecting the most accurate PDF from all constructed joint PDFs.

This section first demonstrates how to introduce DCC into the proposed DMAP estimation. Below, the modified DMAP estimation with DCC is presented in detail.

### A. Introduction of DCC

The introduction of DCC can be divided into three steps.

*1) The choice of the key NWFs*

In the proposed algorithm, exchanging data between neighbors can be viewed as a way of spreading information. The most efficient spreaders are those located within the core of the network as identified by k-shell decomposition analysis [29]; the index of the k-shell is specified as the k-coreness. Nodes with the largest k-coreness consistently a) are infecting larger parts of the network b) are infected more frequently [29]. Therefore, whether the information is transmitted or received, a node with a larger k-coreness value is more comprehensive than a node with a smaller k-coreness value, and the larger one can make a more accurate estimation of the global statistics. Thus the key NWFs are chosen as the nodes with large k-coreness, i.e., the selected key nodes in Fig. 2 are the nodes in pure black with a k-coreness of 3.

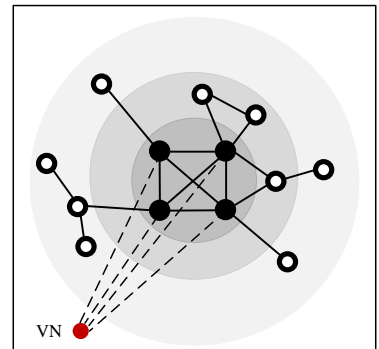

Fig. 2 Schematic diagram of the communication network after the introduction of DCC

*2) The change in topology*

DCC is introduced as a virtual node (VN) linked to the selected key nodes, as shown in Fig. 2. Thereafter, key nodes and VN become neighbors.

*3) The modification of data exchange between neighbors*

All NWFs and VN must estimate the average global statistics to obtain the estimated global statistics. However, VN has no local statistics for estimating its average global statistics. Therefore, we define the estimated average global statistics of VN as the average $\bar{\mathbf{G}}_{j,n}$ from VN's adjacent NWFs, which are the key NWFs. We assume that VN is the (M+1)th nodes in Fig. 2. Let $N_m$ denote the number of NWFs that belong to the adjacent domain $\Omega_m$ of the m-th NWF. Thus, (22) is modified into (24) as follows:

$$\bar{\mathbf{G}}_{j,m}^{t_c+1}=\begin{cases}\bar{\mathbf{G}}_{j,m}^{t_c}+\eta\left[\mathbf{L}_{j,m}-\bar{\mathbf{G}}_{j,m}^{t_c}+\sum_{n\in\Omega_m}\left(\bar{\mathbf{G}}_{j,n}^{t_c}-\bar{\mathbf{G}}_{j,m}^{t_c}\right)\right], & m\leq M\\ \frac{1}{N_m}\sum_{n\in\Omega_m}\bar{\mathbf{G}}_{j,n}^{t_c}, & m=M+1\end{cases} \quad (24)$$

After the converged $\bar{\mathbf{G}}_{j,m}$ is obtained, all NWFs and VN calculate their converged $\mathbf{G}_{j,m}$ via (23). Thus, all NWFs and VN calculate $\mathbf{\Psi}_j$ in (20) independently and complete the first step of the (t+1)th iteration.

The second step of the (t+1)th iteration is the same as the DMAP estimation proposed in Section IV.

Because VN combines the estimated statistics of the key NWFs, after the convergence of the two-step iterative calculation, the (M+1)th joint PDF, which is constructed by VN, can be seen as the choice that best matches the real joint distribution of AWO and FWO.

### B. DMAP estimation with DCC

The modified DMAP estimation with DCC is detailed as follows.

1. Each NWF estimates the aggregated output via (16) and (17)
2. The key NWFs are chosen, and VN is linked to the key NWFs
3. Initialize $\boldsymbol{\varphi}^0$, t=0. The loop is performed until convergence:
   1) Each NWF calculates the local statistics by (19) and obtains $\mathbf{L}_{j,m}^t=\{C_{j,m}^t, \boldsymbol{\beta}_{j,m}^t, \boldsymbol{\gamma}_{j,m}^t, \boldsymbol{\chi}_{j,m}^t, \mathbf{S}_{j,m}^t\}$
   2) Initialize $\bar{\mathbf{G}}_{j,m}^0=0$, $t_c=0$. The loop is performed until convergence:
      a) Each NWF or VN estimates the average global statistics $\bar{\mathbf{G}}_{j,m}^{t_c+1}$ by (24)
      b) $t_c=t_c+1$
   3) Each NWF or VN calculates the converged estimated global statistics $\mathbf{G}_{j,m}^t$ by (23) and then obtains $\mathbf{\Psi}_j^t$ by (20)
   4) Each NWF or VN updates the parameters $\boldsymbol{\varphi}^t=\{w_j^t, \boldsymbol{\mu}_j^t, \boldsymbol{\rho}_j^t \mid j=1,\ldots,J\}$ by (15)
   5) t=t+1
4. Each NWF or VN obtains its optimal $\boldsymbol{\varphi}_{\text{MAP}}$ and then calculates the optimal $\boldsymbol{\theta}_{\text{MAP}}$ via the inverse calculation.
5. The conditional PDF of the aggregated WFE in (7) is derived from the joint PDF constructed by VN via (8).

## VI. CASE STUDY

### A. Data information

The "eastern wind integration data set" of the National Renewable Energy Laboratory (NREL) [30] is used in this paper. This historical wind power data consist of the hourly AWO and their forecast values produced by the Weather Research and Forecasting model [31]. Ten wind farms in Iowa are chosen. Moreover, through a comparison of the estimation in different cases with different choices of key NWFs and the different percentages of k-coreness, the key NWFs are finally chosen as the nodes with top 30% k-coreness. Meanwhile, through experiments with varied GMM order, the number of components is set to 20 for the best estimation effect. Furthermore, after considering both the estimation effect and the construction costs for communication lines, we set the distance threshold for a neighbor node to 4 km. Due to space limitations, details on the case experiments were made available online in [32], which are and will remain freely accessible.

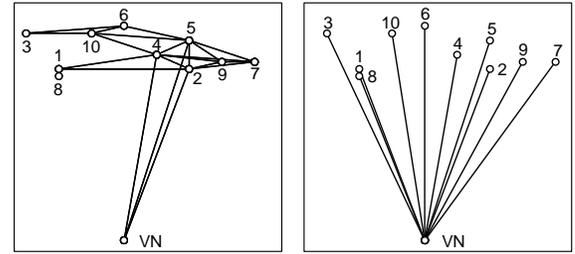

(a) The distributed structure  (b) The centralized structure
Fig. 3 The communication structure of the 10 NWFs and VN

For the proposed algorithm, the corresponding communica-tion network is shown in Fig. 3(a). For the centralized EM algorithm and the centralized MAP estimation, the correspond-ing communication network is shown in Fig. 3(b).

### B. Case 1: Verification of correctness

In this case, a 30-day training data set is used for constructing the joint PDF. The empirical distribution is also formed by the 30-day data records to verify correctness.

The proposed algorithm enables every NWF and VN to construct a joint PDF. To verify correctness of the constructed PDFs, we gather all the estimated aggregated output of all NWFs as the input of the centralized MAP estimation to build a benchmark PDF. Then we utilize root mean squared error (RMSE) [15] to quantify the fitting performance of the proposed algorithm and the centralized MAP estimation. The RMSEs between the marginal PDF of the aggregated AWO constructed by the proposed algorithm and by the centralized MAP estimation are illustrated in descending order in Fig. 4.

In Fig. 4, the RMSEs between the PDF constructed by the proposed algorithm and the benchmark are all lower than $6\times10^{-3}$. In other words, the PDFs constructed by NWFs and VN exhibit little differences from the benchmark, indicating correctness of the proposed algorithm.

Moreover, since the joint PDF is not intuitive, especially when compared to the histogram of the testing data, we give the marginal PDFs derived from joint PDF to show the effect of the proposed algorithm more intuitively. The marginal PDFs of the aggregated AWO and FWO calculated by (6) are shown in Fig. 5(a) and Fig. 5(b), respectively, and the legend 'Proposed algorithm' denotes the results calculated by VN. Fig. 5 shows that the marginal PDFs of the aggregated AWO and FWO obtained by VN both fit the PDFs obtained by the centralized MAP well. Moreover, both values match the empirical distribution. The correctness of the proposed algorithm is verified as well.

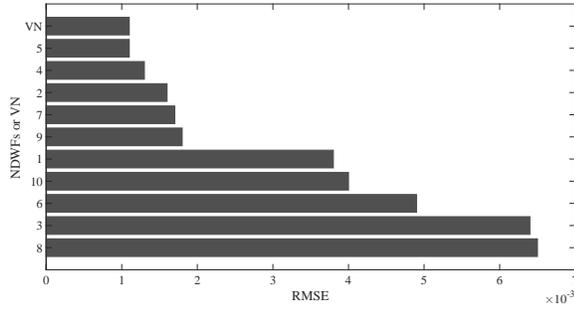

Fig. 4 The RMSEs between the marginal PDF of the aggregated AWO constructed by the proposed algorithm and by the centralized MAP estimation

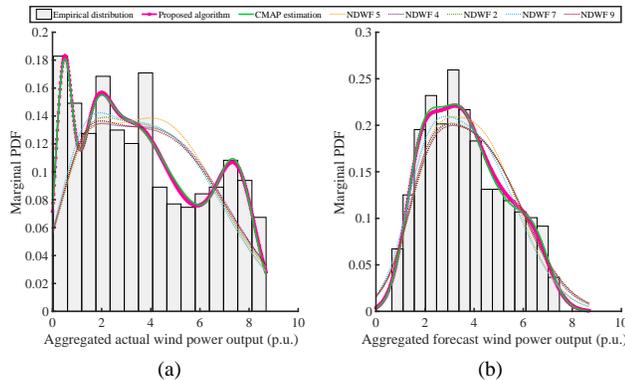

Fig. 5. The marginal PDF of the aggregated actual wind power output in (a) and aggregated forecast wind power output in (b)

Besides, as noted above, once the estimated aggregated output is obtained, each NWF can directly construct the joint PDF of the aggregated AWO and FWO via (14)-(15) by replacing $z_i$ with its own $z_{m,i}$ ($i=1,…, N$). These PDFs are shown in Fig. 5 as dotted lines. For clarity, only 5 PDFs with the smallest RMSEs are shown, which were obtained from NWF 5, NWF 4, NWF 2, NWF 7 and NWF 9. Obviously, these PDFs do not match the empirical distribution. This finding demonstrates the necessity of distributedly training the estimated aggregated output from all NWFs to construct the joint PDF.

Furthermore, according to the above two figures, the PDF calculated by VN has the most desirable fitting performance. Thus, the PDF constructed by VN can be regarded as the best constructed distribution. Therefore, introducing DCC into the DMAP estimation is proven effective for obtaining the most accurate PDF.

*C. Case 2: Verification of the ability to deal with single point failure*

In this case, the 30-day data set is used for training.

After training, 10 NWFs and VN construct 11 joint PDFs with 11 $\theta_{MAP}$ using the proposed algorithm. The $\mu_j$ of the Gaussian components with the highest three weighted coefficients are shown in Fig. 6.

This case aims to verify whether the PDFs can still be constructed after the single point failure of DCC. For the centralized MAP estimation, once DCC is down for any reason, the construction of the PDFs has to be terminated. However, for the proposed algorithm, since the $\mu_j$ calculated by the 10 NWFs and VN are almost the same, indicating that the differences among the 11 joint PDFs are quite small. Thus, every NWF in the distributed communication structure can be considered a backup of DCC. Even though DCC may fail for some reason, the PDFs constructed by other NWFs can be utilized temporarily by the system operator to make optimal decisions. Therefore, the proposed algorithm can deal with the single point failure of DCC.

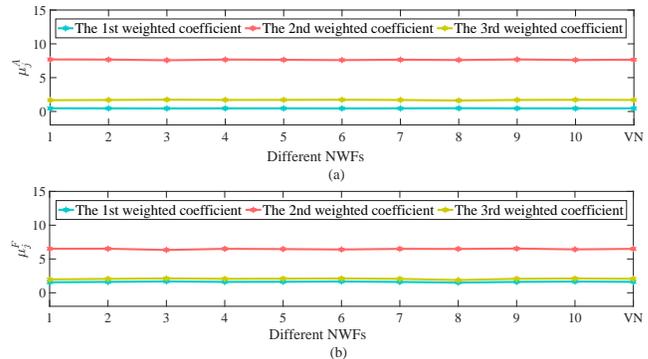

Fig. 6 The $\mu_j$ of the Gaussian components with the highest three weighted coefficients constructed by proposed algorithm: (a) the $\mu_j^A$ of the $\mu_j$; (b) the $\mu_j^F$ of the $\mu_j$.

*D. Case 3: Verification of the robustness*

In practice, long-distance communication is prone to fail. To test the robustness of the proposed algorithm to communication failure, the links between the NWFs are cut off one at a time if the communication distance is over 3.5 km. There are 8 long-distance communication paths: line 1-2, line 1-4, line 2-5, line 2-11, line 4-10, line 4-11, line 5-7, and line 5-11. Similarly, for more intuitive illustration, the marginal PDFs of the aggregated AWO and FWO constructed by VN in different scenarios are given in Fig. 7, where 'Original' represents the intact communication structure and 'Line $i$-$j$' represents the case in which the communication between $i$ and $j$ is cut off. Even though communication failures occur, the PDF constructed by VN exhibits little difference from the original PDF. Therefore, the proposed method has strong robustness to communication failures.

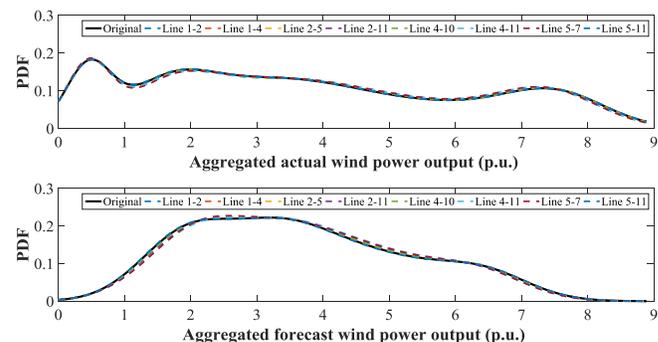

Fig. 7. The marginal PDFs of the aggregated AWO and the aggregated FWO in different scenarios

*E. Case 4: Verification of the ability to solve the limited training data problem via joint PDF*

In this case, 1-day, 2-day, …, and 30-day training data sets are used for training. A 100-day testing data set is used to formulate the marginal empirical distribution for verification. Because the duration of the training data set is no more than 30 days, if the PDFs constructed by VN can fit the empirical distribution well, then the ability to handle the limited training data problem can be verified.

We utilize the proposed algorithm and the centralized EM algorithm to construct the joint PDF. The RMSEs between the empirical marginal distribution and the marginal PDF constructed by the two algorithms are shown in Fig. 8, and the legend 'Proposed algorithm' denotes the results calculated by VN. The length of the training data set is increased from 1 day to 30 days. The RMSEs between the PDF constructed by the centralized EM algorithm and the empirical distribution are extremely high when the amount of training data is small. These results indicate that the centralized EM algorithm will be inaccurate when faced with the limited training data. However, when the training data are limited, the RMSEs between the PDF constructed by the proposed algorithm and the empirical distribution are still approximately 0.1, even though only 1-day data are used for training. With the increasing amount of data for training, the RMSEs of the PDF constructed by the two algorithms both decrease, but the RMSEs from the EM algorithm are always higher than the RMSEs from the proposed algorithm.

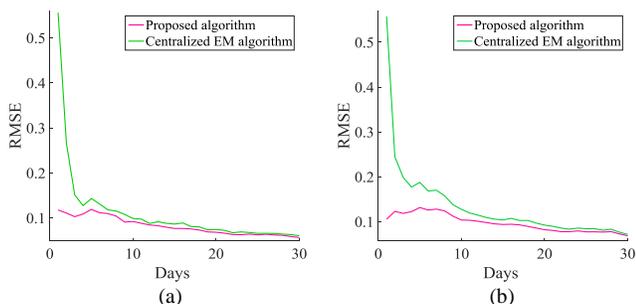

Fig. 8. (a) The RMSEs between the empirical marginal distribution of the aggregated AWO and the corresponding marginal PDF constructed by the two algorithms; (b) the RMSEs between the empirical marginal distribution of the aggregated FWO and the corresponding marginal PDF constructed by the two algorithms

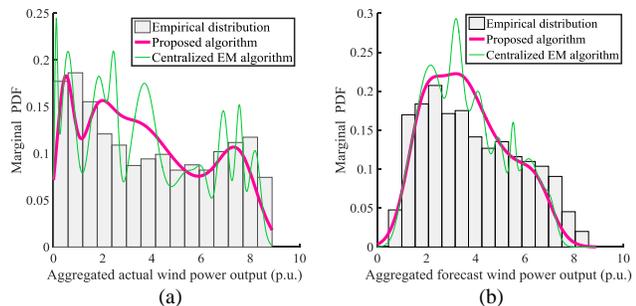

Fig. 9. (a) The marginal PDF of the aggregated AWO and (b) the marginal PDF of the aggregated FWO

The overfitting of the centralized EM algorithm results in high RMSE, especially when the training data are limited. However, through introducing the prior information, the proposed algorithm greatly reduces the degree of the overfitting and provides a smooth PDF curve. For more intuitive illustration, the marginal PDFs calculated by the two algorithms via the 30-day training data set and the marginal empirical distribution constructed via the 100-day testing data set are shown in Fig. 9; the legend 'Proposed algorithm' denotes the calculated results by VN. The curves of the marginal PDFs obtained by the centralized EM algorithm have many sharp peaks, but the marginal PDFs obtained by the proposed algorithm match the empirical distribution well, which means that the proposed algorithm can capture the essential pattern of the whole historical data set from limited historical data.

*F. Case 5: Verification of the ability to solve the limited training data problem via conditional PDF*

In this case, the 1-day, 5-day, 15-day and 30-day training data sets are used for training. The 100-day testing data set is used to formulate the conditional empirical distribution for testing. The conditional empirical distribution is formulated by three steps.

*Step 1*: $N_b$ bins are formulated. Let $y_{max}$ be the maximum of the aggregated FWO value. Each bin has a basic central value $y_c$ and width $a$, where $y_c=0.1 \times y_{max}$ and $a=0.05 \times y_{max}$. Then, the $n_b$th bin is denoted by $[n_b \times y_c - a, n_b \times y_c + a]$.

*Step 2*: 100-day historical data pairs are allocated to $N$ bins. If the aggregated FWO value is in the $n$th bin, then this data pair of the aggregated WFE and FWO is allocated to the $n_b$th bin.

*Step 3*: Aggregated WFEs that belong to the $n$th bin are counted. Then, the histogram of the $n$th bin is formulated based on its aggregated WFE values. This histogram is considered the conditional empirical distribution of the aggregated WFEs for a given forecast value bin. Finally, $N_b$ conditional empirical distributions are obtained.

In this case, $N_b$ is specified to be 9. Compared to the 9 conditional empirical distributions, the RMSE of the conditional PDFs constructed by the proposed algorithm and the centralized EM algorithm are given in Fig. 10, respectively. First, the RMSE from the centralized EM algorithm is extremely high when the training data is limited, but the RMSE from the proposed algorithm is much less due to its ability to overcome the limited training data problem. Second, the RMSE from the proposed algorithm gradually declines with increasing amounts of training data, but some bins in the centralized EM algorithm rise due to the instability caused by overfitting. Third, the RMSE from the centralized EM algorithm is always higher than the RMSE from the proposed algorithm.

The conditional PDFs constructed by the 15-day training data set are shown in Fig. 11. The overfitting of the centralized EM algorithm is still obvious. However, the conditional PDFs constructed by the proposed algorithm match the empirical distribution well with a smooth curve.

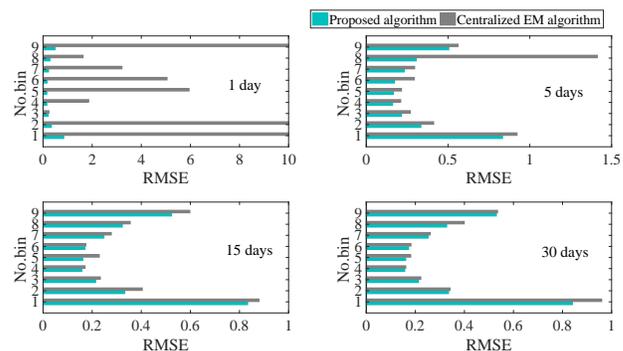

Fig. 10. The RMSE between the conditional PDFs of the aggregated WFE obtained by the two algorithms and the conditional empirical distribution.

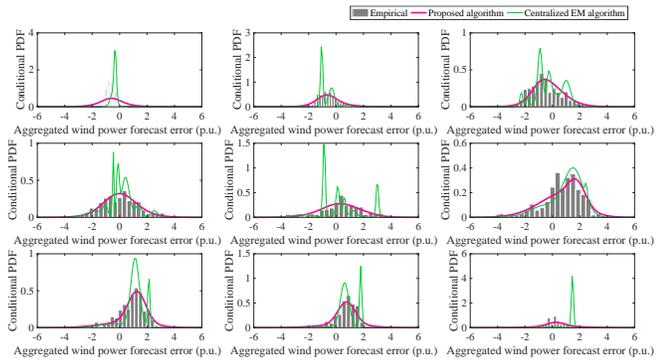

Fig. 11. Conditional PDFs of the aggregated wind power forecast error for 9 bins

## VII. Conclusion

In this paper, the MAP estimation for GMM is applied to address the limited training data issue. Then a DMAP estimation is proposed based on the ACF algorithm to protect data privacy of NWFs with different stakeholders. Finally, DCC is introduced to obtain the best constructed conditional PDF of the aggregated WFE for a given forecast value.

The proposed algorithm can capture the essential pattern of wind power uncertainty by limited training data. In addition, this algorithm protects the data privacy of different NWFs. Moreover, after DCC is introduced as VN, the PDF constructed by VN describes the real distribution best. Furthermore, even if DCC fails, the PDFs built by other NWFs can still be used for decision making, thereby guaranteeing the ability of the proposed algorithm to deal with single point failure of DCC. Besides, the proposed algorithm is robust to communication failures.

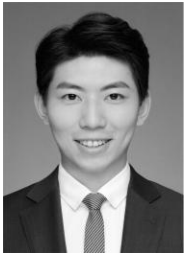

**Mengshuo Jia** (S'18) received his B.E. degree in Electrical Engineering from North China Electric Power University, BaoDing, China, in 2016. He is pursuing his Ph.D. degree in electrical engineering at Tsinghua University, Beijing, China. His research interests include power system probabilistic analysis and renewable energy generation.

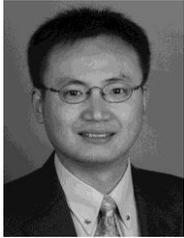

**Chen Shen** (M'98–SM'07) received his B.E. and Ph.D. degrees in Electrical Engineering from Tsinghua University, Beijing, China, in 1993 and 1998, respectively. From 1998 to 2001, he was a postdoc in the Department of Electrical Engineering and Computer Science at University of Missouri Rolla, MO, USA. From 2001 to 2002, he was a senior application developer with ISO New England Inc., MA, USA. He has been a Professor in the Department of Electrical Engineering at Tsinghua University since 2009. Currently, he is the Director of Research Center of Cloud Simulation and Intelligent Decision-making at Energy Internet Research Institute, Tsinghua University. He is the author/coauthor of more than 150 technical papers and 1 book, and holds 21 issued patents. His research interests include power system analysis and control, renewable energy generation and smart grids.

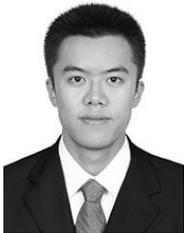

**Zhiwen Wang** (S'13) received his B.E. degree from the School of Electrical Engineering, Beihang University, Beijing, China, in 2013, and his Ph.D. degree in electrical engineering from Tsinghua University, Beijing, China, in 2018. From October 2017 to June 2018, he was a visiting scholar at Washington State University, WA, USA. From June 2018 to October 2018, he was a visiting scholar at Argonne National Laboratory, Argonne, IL, USA. Currently, he works as an engineer at Department of Power Systems, China Electric Power Research Institute.